\documentclass[12pt]{iopart}
\usepackage{iopams}
\usepackage{epsfig}
\usepackage{textcomp}

\begin{document}

\title[Self-organization towards optimally interdependent networks]{Self-organization towards optimally interdependent networks by means of coevolution}

\author{Zhen Wang,$^{1,2}$ Attila Szolnoki,$^{3,4}$ and Matja{\v z} Perc$^{5,*}$}
\address{$^1$Department of Physics, Hong Kong Baptist University, Kowloon Tong, HK\\
$^2$Center for Nonlinear Studies \& the Beijing-Hong Kong-Singapore Joint Center for Nonlinear and Complex Systems, Hong Kong Baptist University, Kowloon Tong, HK\\
$^3$Institute of Technical Physics and Materials Science, Research Centre for Natural Sciences, Hungarian Academy of Sciences, P.O. Box 49, H-1525 Budapest, Hungary\\
$^4$Institute of Mathematics, CNY, P.O. Box 166, H-4400 Ny{\'{\i}}regyh{\'a}za, Hungary\\
$^5$Department of Physics, Faculty of Natural Sciences and Mathematics, University of Maribor, Koro{\v s}ka  cesta 160, SI-2000 Maribor, Slovenia}
\ead{$^{*}$matjaz.perc@uni-mb.si}

\begin{abstract}
Coevolution between strategy and network structure is established as a means to arrive at optimal conditions for resolving social dilemmas. Yet recent research highlights that the interdependence between networks may be just as important as the structure of an individual network. We therefore introduce coevolution of strategy and network interdependence to study whether it can give rise to elevated levels of cooperation in the prisoner's dilemma game. We show that the interdependence between networks self-organizes so as to yield optimal conditions for the evolution of cooperation. Even under extremely adverse conditions cooperators can prevail where on isolated networks they would perish. This is due to the spontaneous emergence of a two-class society, with only the upper class being allowed to control and take advantage of the interdependence. Spatial patterns reveal that cooperators, once arriving to the upper class, are much more competent than defectors in sustaining compact clusters of followers. Indeed, the asymmetric exploitation of interdependence confers to them a strong evolutionary advantage that may resolve even the toughest of social dilemmas.
\end{abstract}

\maketitle

\section{Introduction}
\label{intro}
Networks form the backbone of many complex systems, ranging from human societies to the Internet \cite{pastor_07,barrat_08,christakis_09}, and the structure and function of these networks have been investigated intensely during the last fifteen years \cite{albert_rmp02,newman_siamr03,boccaletti_pr06}. Only recently has the traditional description of complex systems by means of individual networks begun giving way to using multiplex or interdependent networks instead \cite{havlin_pst12}. This is due to the realization that the interdependence between networks might be just as important as the structure of an individual network for the functioning of a system, and because even small and seemingly irrelevant changes in one network can have catastrophic consequences in other networks \cite{buldyrev_n10}. It has also been emphasized time and again that the units constituting a complex system will rarely be connected by the same type of links \cite{kurant_prl06,morris_prl12}, and that thus a single-network description of such systems inevitably entails some loss of information and detail. Using interdependent rather than isolated networks to describe complex systems therefore represents an important step forward towards a more integrative understanding of their dynamics and evolution. Ample efforts have already been invested in the research of cascading failures \cite{buldyrev_n10,li_w_prl12,parshani_pnas11,brummitt_pnas12}, competitive percolation \cite{parshani_prl10,nagler_np11,zhao_jsm13,schneider_pre13}, transport \cite{kurant_prl06,morris_prl12}, diffusion \cite{gomez_prl13}, financial trading \cite{feng_l_pnas12}, neuronal synchronization \cite{sun_xj_chaos11}, as well as statistical mechanics \cite{bianconi_pre13} on interdependent networks. Robustness against attack and assortativity have been studied as well \cite{kurant_pre07,huang_xq_pre11,zhou_d_pre12}, and overall, networks of networks have evolved into a vibrant topic that is of interest to researchers across social and natural sciences \cite{donges_epjb11,gao_jx_np12,helbing_n13,csermely_tde13,halu_epl13}.

Evolutionary games \cite{hofbauer_98,nowak_06} have followed closely the many advances in network science, and it is now thoroughly established that the structure of an interaction network plays a key role by the evolution of cooperation \cite{szabo_pr07,roca_plr09,perc_jrsi13}. While the seminal discovery of network reciprocity is due to Nowak and May \cite{nowak_n92b}, subsequent works have highlighted the importance of small-world \cite{abramson_pre01,tomassini_pre06,fu_epjb07}, scale-free \cite{santos_prl05,santos_pnas06,gomez-gardenes_prl07,rong_pre07,assenza_pre08,santos_n08,pena_pre09,poncela_pre11,brede_epl11,tanimoto_pre12,pinheiro_pone12,simko_pone13}, as well as bipartite \cite{gomez-gardenes_c11,gomez-gardenes_epl11} and coevolving \cite{ebel_pre02,zimmermann_pre04,pacheco_prl06,santos_ploscb06,fu_pa07,tanimoto_pre07,zhang_cy_pone11} networks. Coevolutionary games in particular, where besides strategies the interactions are subject to evolution as well, have increased our understanding of the emergence of system properties that facilitate the evolution of cooperation \cite{perc_bs10}. Most recently, evolutionary games have also been studied on interdependent networks \cite{wang_z_epl12,gomez-gardenes_srep12,gomez-gardenes_pre12,wang_b_jsm12,wang_z_srep13,szolnoki_njp13}, and it has been shown that the interdependence may give rise to new mechanisms that promote the evolution of cooperation. Interdependent network reciprocity \cite{wang_z_srep13}, complex organization of cooperation across different network layers \cite{gomez-gardenes_srep12,gomez-gardenes_pre12}, and information sharing \cite{szolnoki_njp13} are perhaps the most recent examples to demonstrate the case in point.

Emergence of the optimal network interdependence, however, has not yet been studied. More precisely, while the consensus is that the interdependence can promote the evolution of cooperation by means of the many aforementioned mechanisms, and while it has been established that there in fact exists an optimal intermediate level of interdependence that works best in deterring defection \cite{wang_z_srep13b,jiang_ll_srep13}, it is not clear how it could have emerged from initially isolated subsystems. Previous works have simply assumed some level of interdependence be there without considering its origin or mechanisms that might have led to its emergence. To amend this, we here adopt the established concept of coevolution in the realm of the prisoner's dilemma game \cite{maynard_82,axelrod_84}, with the aim of introducing an elementary coevolutionary rule that leads towards the spontaneous emergence of optimal interdependence between two initially completely independent networks. The rule is simple indeed, assuming only that players who are often enough successful at passing their strategy to one of their neighbors, regardless of which strategy it is, are allowed to form an external link to the corresponding player in the other network in order to potentially increase their utility. As we will show, this suffices for the system to self-organize into a two-class society in which the resolution of the prisoner's dilemma is most likely, indicating that even seemingly irrelevant and minute additions to the basic evolutionary process might have led to the intricate and widespread interdependence between networks that we witness today in many social and technological systems.

In the remained of this paper, we first describe in detail the game entailing coevolution between strategy and network interdependence, and then proceed with the presentation of the results. Lastly we summarize the main conclusions and discuss their potential implications.

\section{Model}
\label{model}
To begin with, players populate two independent square lattices of size $L \times L$ with periodic boundary conditions. Each player on site $x$ in network A (up) and on site $x^\prime$ in network B (bottom) is initially connected only with its $k=4$ nearest neighbors and designated either as a cooperator $(s_x=C)$ or defector $(s_x=D)$ with equal probability. The accumulation of payoffs $\pi_x$ and $\pi_{x^\prime}$ on both networks follows the same procedure. Namely, each player plays the game with its four neighbors, whereby two cooperators facing one another acquire $R$, two defectors get $P$, whereas a cooperator receives $S$ if facing a defector who then gains $T$. The prisoner's dilemma game is characterized by the temptation to defect $T=b$, reward for mutual cooperation $R = 1$, and punishment $P$ as well as the sucker's payoff $S$ equaling $0$. Here $1 < b \leq 2$ ensures a proper payoff ranking that captures all relevant aspects of the prisoner's dilemma game \cite{nowak_n92b}, and it also enables a relevant comparison with the many preceding works.

Since the coevolutionary rule (to be introduced below) may allow some players to form an additional external link with the corresponding player from the other network, the utilities used to determine fitness are not simply payoffs obtained from the interactions with the nearest neighbors on each individual network, but rather $U_x = \pi_x + \alpha \pi_{x^\prime}$. The parameter $0 \leq \alpha \leq 1$ determines the strength of external links, increasing the utility of the two connected players the more the larger its value. Naturally, for players that are not linked with their corresponding players in the other network, the utility remains unchanged and is equal to $U_x = \pi_x$.

Following the determination of utilities, the strategy transfer is possible only between nearest neighbors on any given lattice, but never between players residing on different networks even if they are connected with an external link. Accordingly, on network A (and likewise on network B) player $x$ can pass its strategy $s_{x}$ to one of its randomly chosen nearest neighbor $y$ with a probability determined by the Fermi function
\begin{equation}
W(s_{x} \rightarrow s_{y}) = w_x  \frac{1}{1+\exp[(U_{y}-U_{x})/K]}\,\,,
\label{fermi}
\end{equation}
where $K$ quantifies the uncertainty related to the strategy adoption process \cite{szabo_pr07}.

The scaling factor $w_x$ in Eq.~\ref{fermi} is key to the coevolutionary rule, and it also determines how likely players will pass their strategy to one of their neighbors. Traditionally $w_x$ is referred to as the teaching activity of players \cite{szolnoki_epl07,szolnoki_njp08}. Initially all players are assigned a minimal teaching activity $w_{min}=0.01$, but upon each successful strategy pass $w_x$ is enlarged according to $w_x=w_x+\Delta$, where $0< \Delta <1$. Likewise, if the attempted strategy pass is unsuccessful, the teaching activity is reduced by the same value according to $w_x=w_x-\Delta$. To avoid frozen states, all scaling factors are kept between $[w_{min},1]$ at all times. Crucially, only when the teaching activity satisfies $w_x \geq w_{th}$ is player $x$ allowed to have an external link to its corresponding player $x^\prime$ in the other network. If $w_x$ drops below the threshold $w_{th}$, the external link is terminated, which in turn involves the loss of a potential additional payoff for those players who unsuccessfully pass their strategy to the neighbors. The making and braking of the links between the two networks thus serves as individual rewarding and punishment that reflects the evolutionary success within a network.

Simulations are performed by means of a random sequential update, where each player on both equal networks receives a chance to pass its strategy once on average during a full Monte Carlo (MC) step. The trials to pass strategy and to create or delete the external links between the two networks are executed simultaneously, but are independent from each other. System size was varied from $L=200$ to $400$ in order to avoid finite size effects, and the equilibration required up to $10^5$ MC steps. Further simulation details are provided in the figure captions and the Results section, as required. The main question to be answered is whether the simple coevolutionary rule can lead to an optimal interdependence between the two networks, or whether the system will drift towards an extreme state where either none or all possible external links between the corresponding players are established. In what follows, we consider parameters $\Delta$ and $w_{th}$ as crucial to determine the outcome of the coevolutionary game, while parameters $\alpha=0.5$ and $K=0.1$ are kept constant without loss of generality. We have verified that variations of both do not qualitatively change the presented results. The critical temptation to defect $b(K=0.1)=1.0357$ at which cooperators die out on an isolated square lattice is used as the benchmark for the effectiveness of interdependence to promote cooperation passed the boundaries imposed by spatial reciprocity alone.

\section{Results}
\label{results}

\begin{figure}
\centerline{\epsfig{file=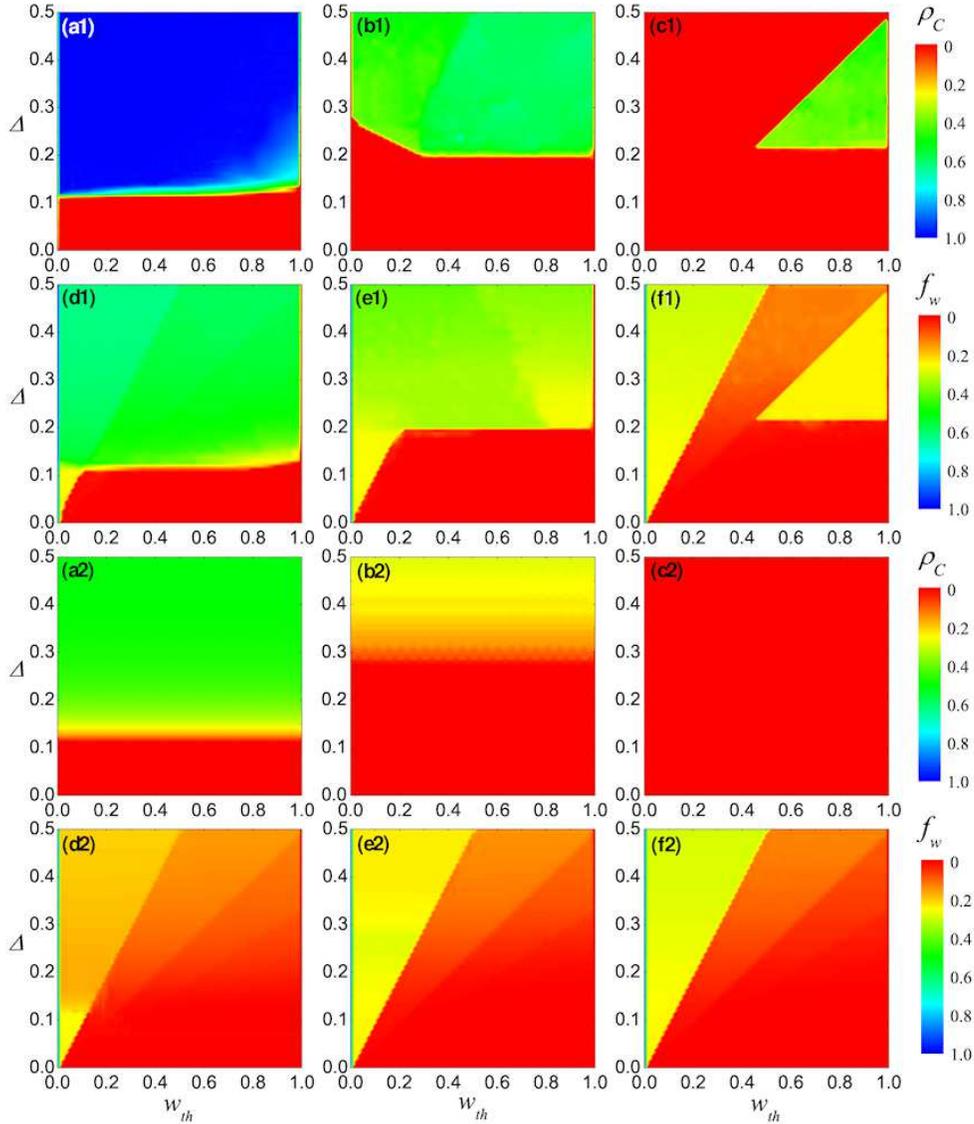,width=13cm}}
\caption{\label{cmaps} Coevolution of cooperation and network interdependence promotes the resolution of the prisoner's dilemma and it allows the interdependence to self-organize towards an optimal level. Panels in the upper row depict the color-encoded fraction of cooperators $\rho_C$ in dependence on the teaching activity threshold to create an external link $w_{th}$ and the reward (punishment) for a successful (unsuccessful) strategy pass $\Delta$ as obtained for $b=1.1$ (a1,d1), $1.29$ (b1,e1) and $1.31$ (c1,f1). The second row depicts the color-encoded fraction of players $f_w$ that fulfill the criterion $w \geq w_{th}$, and are thus allowed to have an external link to the corresponding player in the other network, in dependence on $w_{th}$ and $\Delta$ (for the same values of $b$). The optimal interdependence between the two networks is characterized by approximately half of the players being allowed to have an external link to the other network.
The bottom two rows depict the same results, as obtained in the absence of coupling between the two networks ($\alpha=0$) for $b=1.05$ (a2,d2), $1.1$ (b2,e2) and $1.15$ (c2,f2). The comparison with the results presented in the first two rows clearly demonstrates that the success-dependent teaching activity alone is unable to maintain the high heterogeneity among players that is needed to promote cooperation.}
\end{figure}

Colors maps presented in the top row (a1,b1,c1) of Fig.~\ref{cmaps} encode the fraction of cooperators $\rho_C$ in dependence on the threshold for creating an external link $w_{th}$ and the reward (punishment) for a successful (unsuccessful) strategy pass $\Delta$ that increases (decreases) the teaching activity of the corresponding players. These results provide a comprehensive overview of the impact of the coevolutionary rule on the evolution of cooperation for different values of the temptation to defect $b$. It can be observed that, largely independent of the severity of the social dilemma determined by the value of $b$, there exists a rather sharp threshold in $\Delta$ beyond which cooperative behavior is optimally promoted. The minimally required value of $\Delta$ decreases slightly with decreasing $b$, and for large values of $b$ it also needs to be accompanied by an appropriate intermediate value of $w_{th}$. Actually, for all values of $b$ the threshold of the teaching activity $w_{th}$ should never be neither minimal nor maximal, but rather from within an intermediate interval that becomes narrower as the temptation to defect increases. For comparison, we also show in the third row (a2,b2,c2) of Fig.~\ref{cmaps} the corresponding values of $\rho_C$ that are obtained when the coupling between the two networks is absent ($\alpha=0$). Despite the fact that the value of $w$ for each individual player can still evolve, the success-dependent teaching activity alone is clearly unable to match the high levels of cooperation that we report for the model entailing coevolution of strategy and network interdependence in panels (a1,b1,c1). What is more, the three values of $b$ in panels (a2,b2,c2) are significantly lower than the values of $b$ used in panels (a1,b1,c1), yet still, in decoupled networks, cooperators may altogether fail to survive regardless of $w_{th}$ and $\Delta$ (see c2).

The applied coevolutionary rule thus has the potential to resolve a social dilemma well pass the boundaries imposed by traditional network reciprocity [we remind that the critical temptation to defect at which cooperators die out on an isolated square lattice is $b(K=0.1)=1.0357$] and previously proposed coevolutionary rules affecting only the teaching activity of players. The question is what emergent system property is associated with this observation? Colors maps in the second row (d1,e1,f1) of Fig.~\ref{cmaps} encode the average fraction of players that are allowed to have an external link to the corresponding player in the other network $f_w$, and they provide interesting clues as to a possible answer. Since there is a strong correlation between the fraction of cooperators $\rho_C$ and the fraction of players that have the teaching activity $w \geq w_{th}$, the conclusion that imposes itself is that it is in fact the level of interdependence that emerges between the two networks that plays a key role in warranting an elevated level of cooperation. Only when the self-organization of network interdependence leads to approximately half of the players being allowed to form an external link is the cooperative behavior optimally promoted. This interdependence corresponds rather accurately to the optimum reported in \cite{wang_z_srep13b,jiang_ll_srep13}, only that here it emerges spontaneously from a completely  non-preferential setup with regards to strategy and without any interdependence being assumed in advance.

\begin{figure}
\centerline{\epsfig{file=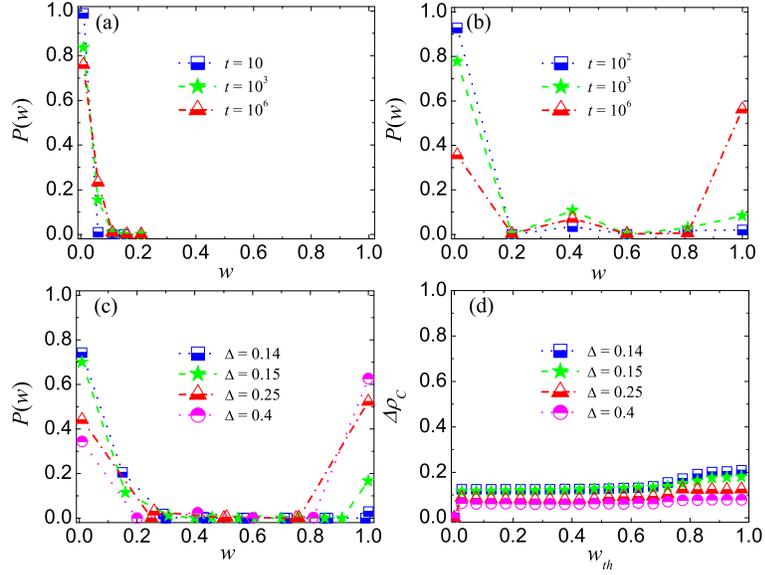,width=10cm}}
\caption{\label{dist} Spontaneous emergence of a two-class society, whereby only the upper class defines the interdependence between the two networks and is able to benefit from it. For low values of the reward (punishment) for a successful (unsuccessful) strategy pass $\Delta$, the initial homogeneous distribution where every player is assigned $w=w_{min}=0.01$ hardly changes towards the stationary state. Even after $10^6$ full MC steps the majority of players still have a marginal teaching activity, and for $\Delta=0.05$ [panel (a)] actually none are able to pass the applied $w_{th}=0.5$ threshold. If, on the other hand, $\Delta$ is above a certain value that we have emphasized in Fig.~\ref{cmaps}, the population segregates spontaneously into players that retain the initially assigned low teaching activity and players that enjoy $w=1$ (and an external link to the corresponding player in the other network since $w \geq w_{th}$). As can be observed in panel (b) ($\Delta=0.4$), the middle class, i.e., players with $w_{min}<w<1$, are practically non-existent as their probability of emergence in the stationary state is $P(w)<0.1$. This spontaneous yet sharp segregation is crucial, as it directly postulates the self-organization towards an optimal network interdependence. The applied temptation to defect in panels (a) and (b) is $b=1.1$. Panel (c) shows the stationary distribution of the teaching activity for four different values of $\Delta$ (see the legend), as obtained for $b=1.15$. If the value of $\Delta$ exceeds $\approx 0.15$, the outcome is always a two-class society, with players belonging either to the upper class and having $w=1$ or to the lower class and having $w=w_{min}$. Panel (d) shows the excess cooperation level in the ``upper class society'' in relation with the average level in the whole system $\Delta \rho_C$ for the same parameters as used in panel (c). If the threshold $w_{th}$ is not too small, the frequency of cooperators in the upper class, i.e., among the players who are able to control their neighborhoods, is always higher than the average.}
\end{figure}

Why precisely the network interdependence self-organizes towards the optimal level and which microscopic system property is responsible can be answered by studying the evolution of the distribution of the teaching activity, as presented in Figs.~\ref{dist}(a) and (b). For $\Delta=0.05$ [panel (a)], which according to the results presented in Fig.~\ref{cmaps} is too low to have a noticeable impact on $\rho_C$ and $f_w$, the vast majority of players retains their initially assigned minimal teaching activity $w=w_{min}$. A minor drift towards $P(w)=0.06$ and $0.11$ can be observed when the stationary state is approached, but importantly, none of the players are able to pass the applied $w_{th}=0.5$ threshold. The two networks therefore remain completely independent, and the evolution of cooperation on both proceeds as on an isolated network. For $\Delta=0.4$ [panel (b)], which is above the minimally required value that can be identified in Fig.~\ref{cmaps}, the outcome is significantly different. The population segregates into a two-class state. Players either have $w=w_{min}$ or $w=1$, but there is very little in-between these two extremes. This is a direct consequence of the reward-punishment mechanism that is utilized in the coevolutionary rule. As soon as a player passes the teaching activity threshold $w_{th}$ to create an external link, its utility is likely to increase further due to the external link to the corresponding player in the other network. This in turn facilitates successful strategy passing, which in turn increases the teaching activity further until the $w=1$ limit is reached (or surpassed, in which case the player is assigned $w=1$). On the other hand, players whose teaching activity does not exceed the $w_{th}$ threshold retain an evolutionary disadvantage, which makes them likelier to fail passing the strategy to their neighbors. This in turn decreases their teaching activity until the $w=w_{min}$ limit is reached (as a technical note, we emphasize that not online the multiples of $\Delta$ plus $w=w_{min}$ are possible, but in fact many other values that emerge if the downgrade begins after the $w=1$ limit is reached). There are of course exceptions to these two most likely outcomes, and even after a player reaches the maximal teaching activity it can still become downgraded to the lower class. Yet such exceptions are rare, and when the stationary state is reached the two-class society is the inevitable outcome if only $\Delta$ is sufficiently large. This conclusion is supported further by the stationary distributions of the teaching activity for several different values of $\Delta$, which are presented in Fig.~\ref{dist}(c). As a consequence, the players in the upper class, and in particular their external links to the other network, give rise to the optimal interdependence between the two networks for the resolution of the prisoner's dilemma. Our argument can be supported further if we compare the result with those obtained where there is no interdependence between the two networks ($\alpha=0$) and the success (or failure) of strategy passing can affect only the teaching activity of players. As the bottom row (d2,e2,f2) of Fig.~\ref{cmaps} shows, in this case the two-class society cannot evolve because the success-dependent teaching activity coevolutionary rule alone is too fragile to maintain the requested heterogeneity among players. Consequently, it is impossible to observe such a significant improvement in the cooperation level as can be observed if the interdependence between the two networks is allowed to emerge and coevolve.

\begin{figure}
\centerline{\epsfig{file=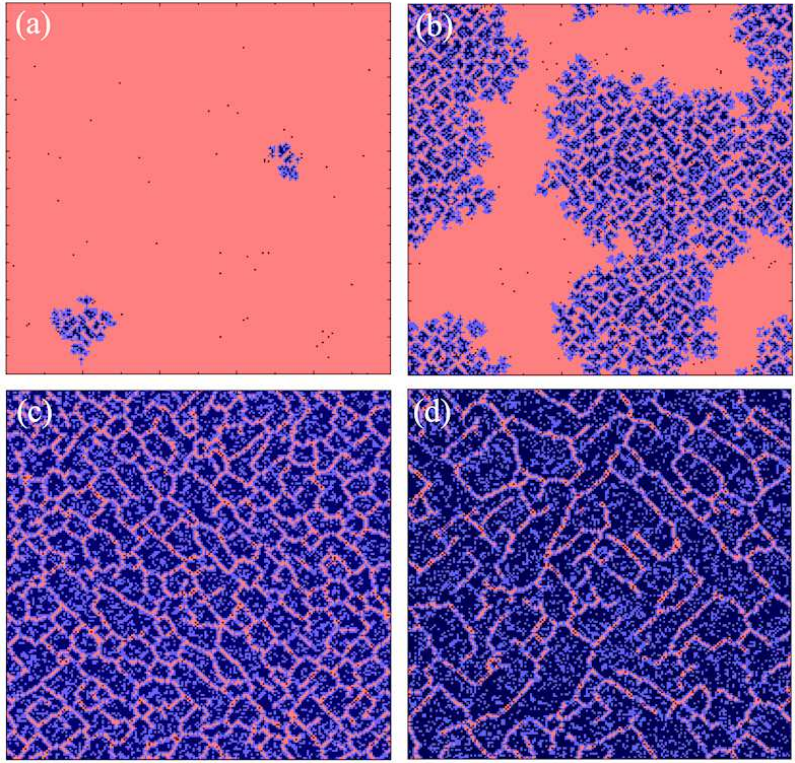,width=10.6cm}}
\caption{\label{snaps} Heterogeneity and interdependent network reciprocity work jointly in favor of the evolution of compact cooperative clusters. Cooperators satisfying $w \geq w_{th}$ and thus belonging to the upper class (depicted dark blue) are able to build up and maintain sizable clusters of followers, although some cooperators there still belong to the lower class (depicted light blue). Upper class defectors (depicted dark red), on the other hand, fail to do likewise. Instead, they exploit their neighbors and suffer the consequences of a negative feedback effect that downgrades them to lower class (depicted light red). Panels (a-d) depict the characteristic stationary distribution of strategies, as obtained for $\Delta=0.14$, $0.15$, $0.25$ and $0.4$, respectively. As soon as $\Delta$ exceeds the required value [panels (c) and (d)], cooperators are able to spread across the whole network. If the applied value of $\Delta$ is too low [panel (a) and to a lesser degree panel (b)], cooperators struggle to survive. These observations are also in full agreement with the corresponding stationary distributions of the teaching activity in Fig.~\ref{dist}(c). The applied temptation to defect is $b=1.15$, the teaching activity threshold for being allowed an external link is $w_{th}=0.8$, and the system size is $L=200$.}
\end{figure}

The consequences of the spontaneous emergence of a two-class society for the evolution of cooperation can be understood best from characteristic snapshots of the square lattice, where the distribution of strategies is colored not only in a strategy-specific way, but also in a $w$-specific way. In Fig.~\ref{snaps}, cooperators (defectors) satisfying $w \geq w_{th}$ and thus having an external link are depicted dark blue (dark red), while cooperators (defectors) failing to meet this criterion are depicted light blue (light red). To facilitate the comparison, the presented snapshots correspond to the stationary distributions of the teaching activity presented in Fig.~\ref{dist}(c). As can be observed, upper class cooperators are masters in building up and maintaining large compact clusters of followers. Defectors, on the other hand, fail to reap long-term benefits of being in the upper class. They simply exploit their neighbors until they themselves become downgraded to the lower class due to a negative feedback effect stemming from the weakening of the neighbors (see also \cite{szolnoki_njp08,szolnoki_epl08} for related work on individual networks). Consequently, the fraction of defectors among successful players is smaller then their frequency in the whole society, as demonstrated by the results presented in Fig.~\ref{dist}(d). The higher the value of $\Delta$, which increases from panel (a) to (d), the more obvious the origin of the evolutionary advantage of cooperators becomes. If $\Delta$ is below a certain value, the segregation of players is either completely absent or marginal [see $\Delta=0.14$ and $0.15$ curves in Fig.~\ref{dist}(c) and Fig.~\ref{dist}(a)], with the majority of players still retaining their initially assigned minimal teaching activity $w=w_{min}$ even in the stationary state. Consequently, cooperators can take advantage neither from a heterogeneous state nor from the interdependence between the two networks \cite{wang_z_srep13}. They therefore remain bounded to isolated and relatively small [especially in panel (a) and to a lesser degree also in panel (b) of Fig.~\ref{snaps}] regions of the network where some individuals nevertheless do succeed in fulfilling $w \geq w_{th}$ (note that for $\Delta=0.14$ and $0.15$ a non-zero fraction of players does satisfy $w=1$). But once $\Delta$ exceeds a threshold (see Fig.~\ref{cmaps} for details), the outcome is always a strongly segregated two-class society [see $\Delta=0.25$ and $0.4$ curves in Fig.~\ref{dist}(c) and Fig.~\ref{dist}(b)], which results in widespread cooperation, as depicted in panels (c) and (d) of Fig.~\ref{snaps}.

\begin{figure}
\centerline{\epsfig{file=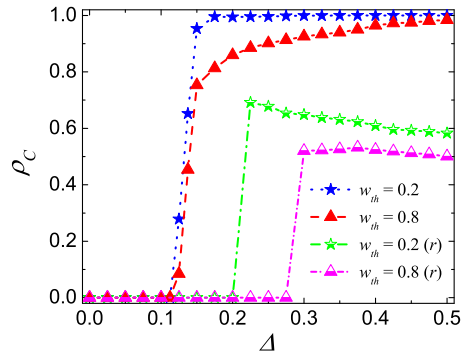,width=6cm}}
\caption{\label{random} With disrupted interdependent network reciprocity the evolution of cooperation is impaired. Depicted is the fraction of cooperators $\rho_C$ in dependence on the reward (punishment) for a successful (unsuccessful) strategy pass $\Delta$. If players satisfying $w \geq w_{th}$ form external links with randomly chosen players from the other network [marked (r) in the figure legend], the rise of $\rho_C$ as $\Delta$ increases past the critical value is lesser than if players form external links with the corresponding players from the other network. Interestingly, the threshold also shifts towards larger values of $\Delta$, indicating that a stronger feedback in the teaching activity is needed to evoke the same effect as when the interdependent network reciprocity can take full effect. The applied temptation to defect is $b=1.12$.}
\end{figure}

While the enhanced clustering of cooperators that sets in for sufficiently large $\Delta$ is akin to traditional network reciprocity \cite{nowak_n92b}, and in particular also to the mechanism that is responsible for elevated levels of cooperation in highly heterogeneous environments \cite{perc_pre08,santos_n08,perc_njp11,santos_jtb12}, the self-organization of interdependence adds another layer of support to the evolution of cooperation. To demonstrate this fact, we study an alternative model where players satisfying $w \geq w_{th}$ do not form an external link with the corresponding player from the other network, but rather with a randomly chosen player. This alteration ought to disrupt interdependent network reciprocity \cite{wang_z_srep13}, and as shown in Fig.~\ref{random} [see label (r)] indeed it does. The jump upwards in $\rho_C$ is significantly lower, and it occurs at higher values of $\Delta$, thus further corroborating the fact that the spontaneous emergence of optimal network interdependence is crucial for the highly cooperative outcome of the prisoner's dilemma game.

\section{Discussion}
\label{discussion}
We have studied the coevolution of strategy and network interdependence with the aim of establishing whether a simple coevolutionary rule can lead to the spontaneous emergence of optimal conditions for the resolution of a social dilemma. Players were allowed to form an external link with the corresponding player in the other network only if they succeeded in maintaining their teaching activity above a certain threshold $w_{th}$. The teaching activity was increased upon each successful strategy pass by $\Delta$, and decreased by the same amount whenever an attempted strategy pass failed. We have shown that due to coevolutionary rule the interdependence between networks self-organizes so that approximately half of the players form an external link, while the other half is denied participation in activities beyond their host network. This level of interdependence corresponds accurately to the manually identified optimum for deterring defection \cite{wang_z_srep13b,jiang_ll_srep13}, with the important difference that here it emerges spontaneously from two initially completely independent networks.

In terms of the outcome of the prisoner's dilemma, we have assumed that players that do have an external link may benefit from it by means of an increased utility, but are not allowed to pass strategy across it. We have observed the spontaneous emergence of a two-class society, where the upper class defined the interdependence between the two networks and was able to enjoy the benefits of high teaching activity and increased utility, while the lower class had to make do with low potency and lack of benefits stemming from the interdependence. Remarkably, the middle class was found to be practically non-existent. Despite the fact that the coevolutionary rule is strategy independent, the spatial patterns have revealed that cooperators forge evolutionary advantages from the upper class status much more effectively than defectors. While the latter simply exploit their neighbors until they themselves become downgraded to lower class (assuming they manage to get into the upper class to begin with), cooperators work together to form compact clusters on both networks that further reinforce themselves via interdependent network reciprocity \cite{wang_z_srep13}. As our simulations have revealed, the cooperation level in the upper class is actually always higher than in the whole population. For the coevolutionary rule to work optimally in favor of the evolution of cooperation, $\Delta$ has to be sufficiently large, i.e., comparable with the actual teaching activity, while the threshold $w_{th}$ has to be neither too small nor too large. These conditions ensure that the reward and punishment related to the successful and unsuccessful strategy pass are sufficiently notable to take a rapid enough effect, as otherwise the feedback from the teaching activity and the interdependence might be either too weak or too slow to notably affect the resolution of the social dilemma. It is also important to emphasize that the coevolving teaching activity alone is unable to produce the optimal two-class society within a network. The possibility to establish an external link -- the coevolution of interdependence between the two networks -- is thus an essential ingredient that is needed for the segregated society to emerge. With these new insights, it is now possible to understand that the previously established optimal strength of interdependence between networks implicitly generated the required heterogeneity within a network for the optimal evolution of cooperation. We have also shown that, if players are allowed to form external links with randomly chosen rather than corresponding players from the other network, the evolution of cooperation is significantly impaired, thus additionally highlighting the importance of interdependent network reciprocity.

Presented results strongly corroborate the mounting evidence that the interdependence between networks may be just as important as the structure of an individual network \cite{buldyrev_n10,havlin_pst12,gao_jx_np12,helbing_n13}. We have studied the effects of the coevolution of teaching activity on an isolated network before \cite{szolnoki_njp08}, and at that time we have shown that it leads to the spontaneous emergence of highly heterogeneous states that promote the evolution of cooperation \cite{perc_pre08,santos_n08,perc_njp11,santos_jtb12}. Here, the coevolution affects also the interdependence between the two square lattices, and the positive effect on the evolution of cooperation is indeed much stronger. In \cite{szolnoki_njp08} we were unable to observe cooperation for $b>1.18$, while at present cooperators are able to match defectors in the occupancy of the population even for $b>1.3$. Under the same conditions the critical temptation to defect on an isolated square lattice is $b(K=0.1)=1.0357$, so that thus both coevolutionary rules markedly improve the survival chances of cooperative behavior, yet building also on the interdependence between the two networks is significantly more effective still.

Awarding success is commonplace in human societies, as is punishing failure \cite{sigmund_10}. In this regard, the proposed coevolutionary rule captures an elementary ingredient of real-life interactions, and in doing so demonstrates that this alone is sufficient to arrive at a healthy level of interdependence between two initially independent populations. We hope this will inspire more research aimed at understanding how the intricate interdependence between many very different networks came to be. Taking into account the aspect of growth as an ever present ingredient of individual networks \cite{poncela_ploso08,poncela_njp09,portillo_pre12} might be particularly inviting, as this would enable a simultaneous study of the emergence of network structure as well as network interdependence. While the former is firmly established as an important factor for the outcome of evolutionary games, and in general the subject is also thoroughly studied and well understood \cite{szabo_pr07,roca_plr09,perc_jrsi13}, disentangling the role of network interdependence certainly merits further attention.

\ack
This research was supported by the Hungarian National Research Fund (Grant K-101490), TAMOP-4.2.2.A-11/1/KONV-2012-0051, and the Slovenian Research Agency (Grant J1-4055).

\section*{References}
\providecommand{\newblock}{}

\end{document}